\documentclass[aps,prl,reprint,superscriptaddress]{revtex4-1}
\usepackage{amssymb}
\usepackage{amsmath}
\usepackage{psfrag}

\newcommand{\iu}{\mathrm{i}}	

\newcommand{\bra}[1]{\ensuremath{\left\langle #1 \right|}}
\newcommand{\ket}[1]{\ensuremath{\left| #1\right\rangle}}
\newcommand{\braket}[2]{\ensuremath{\left\langle #1 | #2\right\rangle}}
\newcommand{\melement}[3]{\ensuremath{\left\langle #1\left| #2 \right| #3\right\rangle}}

\begin{document}

\title{Dressed bound states for attosecond dynamics in strong laser fields}

\author{V. S. Yakovlev}
\email{vladislav.yakovlev@lmu.de}
\affiliation{Ludwig-Maximilians-Universit\"at, Am Coulombwall 1, 85748 Garching, Germany}
\affiliation{Max-Planck-Institut f\"ur Quantenoptik, Hans-Kopfermann-Stra{\ss}e 1, 85748 Garching, Germany}

\author{M. Korbman}
\affiliation{Max-Planck-Institut f\"ur Quantenoptik, Hans-Kopfermann-Stra{\ss}e 1, 85748 Garching, Germany}

\author{A. Scrinzi}
\affiliation{Ludwig-Maximilians-Universit\"at, Theresienstra{\ss}e 37, 80333 Munich, Germany}

\begin{abstract}
  We propose a theoretical approach for the interpretation of
  pump-probe measurements where an attosecond pulse is absorbed in the
  presence of an intense laser pulse. This approach is based on
  abstractly defined \emph{dressed bound states}, which capture the
  essential aspects of the interaction with the laser pulse and
  facilitate a perturbative description of transitions induced by the
  attosecond pulse. Necessary properties of dressed bound states are
  defined and various choices are discussed and compared to accurate
  numerical solutions of the time-dependent Schr\"odinger equation.
\end{abstract}

\maketitle

\section{Introduction}
\label{introduction}
The ionization of atoms and molecules in the field of an intense laser
pulse is one of the most fundamental phenomena in the interaction of
light with matter. Despite decades of intensive experimental and
theoretical research, there are still open conceptual questions,
highlighted by recent experiments
\cite{Wirth_Science_2011,Uiberacker_Nature_2007}.  A key question is
whether one can define the instantaneous ionization probability in the
presence of an ionizing field.
A closely related question is to
which extent the detachment of an electron by multiphoton absorption
or quantum tunneling can be described by an ionization rate
(see, for example, \cite{Saenz_PRA_2007,Vafaee_PRA_2007}). Such a description 
is common, but it can be made rigorous only in stationary or quasi-stationary situations.
Two such cases are the static and time-periodic external electric fields. 
In static fields,
the system decays exponentially at a well-defined rate.
In time-periodic fields, Floquet theory allows defining
cycle-averaged decay rates (see e.g. \cite{Chu_ACP_1989} or
\cite{vanderhart06_floquet} for a recent application). 
In both situations, real and imaginary 
parts of complex energies give energy positions and decay widths. 
Non-Hilbert space resonance states can be associated with the complex energies,
either by using complex scaling \cite{scrinzi00_ionization_rates,scrinzi93_stabilization} 
or by iteratively adapting the boundary conditions (Siegert boundary conditions) \cite{vanderhart06_floquet}. None of these
approaches, however, provides a fully satisfactory framework for
investigating the interaction of few- to single-cycle ionizing pulses
with matter. At the same time, the establishment of attosecond
absorption spectroscopy
\cite{Pfeifer_CPL_2008,Goulielmakis_Nature_2010} and the recent
advances in the generation of optical waveforms
\cite{Wirth_Science_2011,Chan_Science_2011} call for a deeper
understanding of dynamics unfolding within a laser cycle. Such
understanding implies the development of approximate analytical
theories, and the aim of this article is to make a step towards
creating a useful theoretical framework.

Our work is closely related to the development of adiabatic or
quasistatic approximations to strong-field ionization, such as
described in \cite{Bondar_PRA_2009} and the papers cited therein.  In
contrast to that research, we make no attempt to analytically describe
strong-field ionization. Instead, we search for an analytical
description of resonant transitions that an attosecond pulse of
ultraviolet (UV) or extreme-ultraviolet (XUV) radiation drives in
the presence of a strong near-infrared field.

\section{Dressed bound states}
\label{general-formalism}
Let us consider a quantum system (atom, ion, molecule etc.) described by
a Hamiltonian $\hat{H}_0$. In the absence of any external field, this system has
a set of orthonormal bound stationary states that satisfy the
stationary Schr\"odinger equation:
\begin{equation}
  \label{eq:TISE}
  \hat{H}_0 \ket{n} = \epsilon_n^{(0)} \ket{n}.
\end{equation}
A typical attosecond measurement consists in letting the system
interact with the electric fields of an intense laser pulse
$\mathbf{E}_\mathrm{L}(t)$, which may be regarded as a \emph{pump}
pulse, and a relatively weak probe pulse (or a train of pulses)
$\mathbf{E}_\mathrm{probe}(t)$. An XUV probe pulse may be as short as
a few tens of attoseconds
\cite{Goulielmakis_Science_2008,Mashiko_PRL_2008}. Varying the delay
between the two pulses with an attosecond accuracy and observing the
outcomes of their interaction with a quantum system provides valuable
insights into phenomena triggered and steered by these two pulses
\cite{Krausz_RMP_2009}. If a significant fraction of atoms or
molecules is ionized within a single half-cycle of the laser pulse,
the interaction with the pulse is highly non-perturbative, so that an
accurate description of this interaction calls for a numerical
solution of the time-dependent Schr\"odinger equation (TDSE). We
assume that such a solution $\ket{\psi_\mathrm{L}(t)}$ is available,
and one of our goals is to \emph{interpret} strong-field dynamics in terms of some
dressed states. These states, if introduced properly, may serve as a
convenient starting point for building approximate models describing
the interaction with the probe pulse. Before we discuss different
approaches to defining such dressed states, let us list a few
requirements that they should fulfill.

Let $\ket{\varphi_n(t)}$ be a bound dressed state associated with a
stationary state $\ket{n}$. This means that $\ket{\varphi_n(t_0)} =
\ket{n}$ at some moment $t_0$ before the interaction with external
fields. As a first requirement,
we demand that dressed states should form an orthonormal set
of wave vectors at all times:
\begin{equation}
  \label{eq:orthonormality}
  \braket{\varphi_m(t)}{\varphi_n(t)} = \delta_{m n}.
\end{equation}
Orthonormality is not only convenient for computations, but it is also
needed for the physical interpretation of the states: only for
orthonormal states does finding a system in state $|\varphi_n(t)\rangle$
with probability $|\langle \varphi_n(t)|\psi_\mathrm{L}(t)\rangle|=1$
imply that all other probabilities are 0.


A strong laser field can significantly ionize a quantum system. As a
free electron cannot absorb a photon, the probe pulse predominantly
interacts with bound electrons. Therefore, it would be desirable to
have dressed states that serve as a basis for describing the bound
part of an electron wavefunction even in the presence of a strong
external electric field. When this property is satisfied,
strong-field ionization should deplete bound dressed states without changing
their shape in a fashion similar to the decay of quasi-stationary
states.

Except for losses due to ionization, each dressed state should fully
account for distortions caused by the laser field. Therefore, we
assume that the pump pulse induces no transitions between dressed
states. This is the most important requirement that we place on
dressed states, and it means that
$\melement{\varphi_m(t_2)}{\hat{U}_\mathrm{L}(t_2,t_1)}{\varphi_n(t_1)}$
must be negligibly small if $m \ne n$, where
$\hat{U}_\mathrm{L}(t_2,t_1)$ is the propagator describing the
interaction with the laser pulse (the exact definition of the
propagator is given below).  In particular, we expect that the
interaction with the laser pulse brings the quantum system from its
initial state $\ket{n}$ to the corresponding dressed state
$\ket{\varphi_n(t)}$ multiplied with a probability amplitude $a_n(t)$,
the modulus of which accounts for the depletion
by the laser pulse:
\begin{equation}
  \label{eq:probability.amplitude}
  a_n(t) = \melement{\varphi_n(t)}{\hat{U}_\mathrm{L}(t,t_0)}{n},
\end{equation}
while
\begin{equation}
  \melement{\varphi_m(t)}{\hat{U}_\mathrm{L}(t,t_0)}{n} \approx 0
  \ \mbox{if}\ m \ne n.
\end{equation}
As long as the above assumptions hold, the effect of the laser
field on any dressed bound state can be expressed via the
probability amplitudes:
\begin{equation}
  \label{eq:propagator_matrix}
  \melement{\varphi_n(t_2)}{\hat{U}_\mathrm{L}(t_2,t_1)}{\varphi_n(t_1)} \approx
  \frac{a_n(t_2)}{a_n(t_1)},
\end{equation}
and the time evolution of the system can be approximated as 
\begin{equation}
\hat{U}_\mathrm{L}(t_2,t_1)\approx\sum_n|\varphi_n(t_2)\rangle \frac{a_n(t_2)}{a_n(t_1)}\langle\varphi_n(t_1)|.
\end{equation}
Note that the approximate time evolution is no longer unitary as it accounts for ionization by losses from the 
dressed-state space.

Assuming that time-dependent wave vectors $\ket{\varphi_n(t)}$
satisfying all the above requirements exist, let us use them to
develop a general perturbative description of the interaction of a
laser-dressed system with a weak high-frequency probe pulse.
For simplicity, we assume that
the pump (laser) and probe (UV or XUV) pulses are polarized along the $z$-axis. In the
dipole approximation, the length-gauge Hamiltonian takes the form
$\hat{H}(t) = \hat{H}_0 + \hat{V}_\mathrm{L}(t) +
\hat{V}_\mathrm{probe}(t)$ with
\begin{align}
  \label{eq:external-potential}
  \hat{V}_\mathrm{L}(t) &= e E_\mathrm{L}(t) \hat{Z},\\
  \hat{V}_\mathrm{probe}(t) &= e E_\mathrm{probe}(t) \hat{Z},
\end{align}
where $e>0$ is the electron charge and $\hat{Z}$ is the dipole operator.
The interaction with the laser pulse is described by
\begin{equation}
  \label{eq:TDSE-laser}
  \iu\hbar \frac{d}{dt}\ket{\psi_\mathrm{L}(t)} = \bigl(\hat{H}_0(t) +
  \hat{V}_\mathrm{L}(t) \bigr) \ket{\psi_\mathrm{L}(t)}
\end{equation}
or, in the operator form,
\begin{equation}
  \label{eq:U_L}
  \iu\hbar \frac{\partial}{\partial t} \hat{U}_\mathrm{L}(t,t_0) =
  \bigl(\hat{H}_0 + \hat{V}_\mathrm{L}(t) \bigr) \hat{U}_\mathrm{L}(t,t_0)
\end{equation}
with $\ket{\psi_\mathrm{L}(t)} = \hat{U}_\mathrm{L}(t,t_0)\ket{i}$ for an initial state $\ket{i}$.
A standard approach to developing a perturbation theory with respect
to the probe pulse consists in writing the complete TDSE
\begin{equation}
  \label{eq:TDSE}
  \iu\hbar \frac{d}{dt}\ket{\psi(t)} = \hat{H}(t) \ket{\psi(t)}
\end{equation}
in the following integral form:
\begin{equation}
  \label{eq:TDSE-integral}
  \hat{U}(t,t_0) = \hat{U}_\mathrm{L}(t,t_0) -
  \frac{\iu}{\hbar} \int_{t_0}^t dt'\,
  \hat{U}(t,t') \hat{V}_\mathrm{probe}(t') \hat{U}_\mathrm{L}(t',t_0).
\end{equation}
Here, $\hat{U}(t,t_0)$ is the propagator associated with the full Hamiltonian
$\hat{H}(t)$:
\begin{equation}
  \label{eq:U}
  \iu\hbar \frac{\partial}{\partial t} \hat{U}(t,t_0) = \hat{H}(t) \hat{U}(t,t_0)
\end{equation}
and $\ket{\psi(t)} = \hat{U}(t,t_0)\ket{i}$ for the initial condition
$\ket{\psi(t_0)} = \ket{i}$.  Furthermore, if only single-photon
processes play a role in the interaction with the probe pulse, then
the unknown operator $\hat{U}(t,t')$ on the right-hand side of
Eq.~\eqref{eq:TDSE-integral} may be approximated with
$\hat{U}_\mathrm{L}(t,t')$, which we consider to be known. This yields
\begin{equation}
  \label{eq:FOP1}
  \ket{\psi(t)} \approx \ket{\psi_\mathrm{L}(t)} -
  \frac{\iu}{\hbar} \int_{t_0}^t dt'\,
  \hat{U}_\mathrm{L}(t,t') \hat{V}_\mathrm{probe}(t') \ket{\psi_\mathrm{L}(t')}.
\end{equation}
The probe pulse may cause bound-bound, as well as bound-continuum
transitions. In this paper, we focus on transitions between bound
states, leaving the direct photoionization by the probe pulse aside. In this
case, Eq.~\eqref{eq:FOP1} can be written as
\begin{multline}
  \label{eq:FOP2}
  \ket{\psi(t)} \approx \ket{\psi_\mathrm{L}(t)} -
  \frac{\iu}{\hbar} \sum_{l,m} \int_{t_0}^t dt'\,
  \Biggl\{
  \hat{U}_\mathrm{L}(t,t') \ket{\varphi_l(t')} \times \\
  \melement{\varphi_l(t')}{\hat{V}_\mathrm{probe}(t')}{\varphi_m(t')}
  \braket{\varphi_m(t')}{\psi_\mathrm{L}(t')} \Biggr\}.
\end{multline}

Let us evaluate the probability amplitude that a system, initially prepared in
state $\ket{\psi(t_0)}=\ket{i}$, will be found in state
$\ket{\varphi_{n \ne i}(t)}$ at a later time $t$. Multiplying both sides of
Eq.~\eqref{eq:FOP2} with $\bra{\varphi_n(t)}$ from the left,
keeping in mind the fact that $\ket{\psi_\mathrm{L}(t)}
\approx a_i(t) \ket{\varphi_i(t)}$, and using
Eq.~\eqref{eq:propagator_matrix}, we obtain
\begin{multline}
  \label{eq:FOP3}
  \alpha_{n i}(t) = \melement{\varphi_n(t)}{\hat{U}(t,t_0)}{i} =
  \braket{\varphi_n(t)}{\psi(t)} \approx\\
  - \frac{\iu}{\hbar} e a_n(t)
  \int_{t_0}^t dt'\, E_\mathrm{probe}(t')
  \frac{a_i(t')}{a_n(t')} Z_{n i}(t')
\end{multline}
for $n \ne i$, where
\begin{equation}
  \label{eq:dipole}
  Z_{n i}(t) = \melement{\varphi_n(t)}{\hat{Z}}{\varphi_i(t)}
\end{equation}
is the transition matrix element between dressed
states. Eq.~\eqref{eq:FOP3} is the desired expression, where the
interaction with the probe pulse is accounted for in first-order
perturbation theory, while the effect of the strong laser pulse is
represented by quantities related to dressed states. Based on
Eq.~\eqref{eq:FOP3}, we can evaluate quantities observable in
measurements. For example, the probability that, at some final time
$t_\mathrm{f}$ after the interaction, the system will be found in a
bound stationary state $\ket{f}$ is given by
\begin{equation}
  \label{eq:transition.probability}
  p_{f i} = \left|  \braket{f}{\psi(t_\mathrm{f})} \right|^2 =
  \left| \sum_n \alpha_{n i}(t_\mathrm{f}) \braket{f}{\varphi_n(t_\mathrm{f})} \right|^2.
\end{equation}
As an example that is more relevant to transient absorption
spectroscopy, the dipole response due to bound-bound
transitions induced by the probe pulse can be evaluated as
\begin{equation}
  \label{eq:dipole-response}
  d(t) = \melement{\psi(t)}{\hat{Z}}{\psi(t)} \approx
  \sum_{m,n} \alpha_{m i}^*(t) Z_{m n}(t) \alpha_{n i}(t).
\end{equation}

So far, the dressed bound states were treated as an abstract
mathematical concept. It is worth mentioning that without the
requirement that dressed states should describe bound electrons, the
wavevectors $\ket{\varphi_n(t)}=\hat{U}_\mathrm{L}(t,t_0)\ket{n}$
would exactly satisfy the rest of our requirements, but such
dressed states would not serve their main purpose, which is to
facilitate the interpretation of bound electron dynamics in the
presence of an intense laser pulse. We do not know if perfect dressed
states satisfying all our requirements including their bound character exist,
but we can consider several approximations to dressed bound states and
use our general theory to investigate their performance in comparison
with exact solutions of the TDSE. We will consider four different
definitions of dressed bound states.

As a very crude approximation, one can attempt to use the unperturbed
stationary states:
\begin{equation}
  \label{eq:varphi0}
  \ket{\varphi_n^{(u)}(t)} = \ket{n}.
\end{equation}
Another option, frequently encountered in literature, is to use
time-dependent eigenstates of the Hamiltonian $\hat{H}(t)$ evaluated
in the subspace of unperturbed bound states and normalized to satisfy
Eq.~\eqref{eq:orthonormality}:
\begin{equation}
  \label{eq:varphi1}
  \sum_m \ket{m}
  \melement{m}{\hat{H}_0 + \hat{V}_\mathrm{L}(t)}{\varphi_n^{(a)}(t)} =
  \epsilon_n^{(a)}(t) \ket{\varphi_n^{(a)}(t)}.
\end{equation}
In the context of strong-field ionization, these states are known
under many different names, such as adiabatic states
\cite{Bondar_PRA_2009,Staudte_PRL_2007}, quasistatic states
\cite{Dietrich_PRL_1996,Kelkensberg_PCCP_2011}, phase-adiabatic states
\cite{Kawata_JCP_1999}, field-adapted states \cite{Remacle_PRL_2011},
and adiabatic field-following dressed states
\cite{Mies_1993,Harumiya_PRA_2002}. We will refer to them as
``adiabatic states''.

Yet another reasonable approach is to obtain dressed bound states
by solving the TDSE in the basis of unperturbed bound states:
\begin{equation}
  \label{eq:varphi2}
  \iu \hbar \frac{d}{dt} \ket{\varphi_n^{(d)}(t)} =
  \sum_n \ket{n}
  \melement{n}{\hat{H}_0 + \hat{V}_\mathrm{L}(t)}{\varphi_n^{(d)}(t)}
\end{equation}
with the initial condition $\ket{\varphi_n^{(d)}(t_0)}=\ket{n}$.
With this choice, we assume that virtual excitations 
into field-free  continuum states do not influence the dynamics.
However, contrary to $\ket{\varphi_n^{(a)}(t)}$,
these states do not follow the laser field adiabatically, but
rather allow for non-instantaneous response to
the laser field. We will call them ``dynamically dressed states''.

The above
three alternatives allow one to evaluate the dressed bound states without
solving the TDSE in the full Hilbert space, although such solutions
$\ket{\psi_\mathrm{L}(t)}$ are required later to obtain the
probability amplitudes $a_n(t)$ from Eq.~\eqref{eq:probability.amplitude}. The last
option that we are going to consider is different: we will evaluate
dressed bound states by projecting solutions of the TDSE
obtained in the whole Hilbert space \eqref{eq:TDSE-laser} onto the
subspace of unperturbed bound stationary states and orthonormalizing
them with the aid of the Gram-Schmidt procedure:
\begin{align}
  \label{eq:varphi3-tilde}
  \ket{\tilde{\varphi}_n^{(p)}(t)} &=
  \prod_{l<n} \left(1-\hat{P}_l(t)\right)
  \sum_m \ket{m}\melement{m}{\hat{U}_\mathrm{L}(t,t_0)}{n},\\
  \label{eq:varphi3}
  \ket{\varphi_n^{(p)}(t)} &= \ket{\tilde{\varphi}_n^{(p)}(t)} /
  \sqrt{\braket{\tilde{\varphi}_n^{(p)}(t)}{\tilde{\varphi}_n^{(p)}(t)}},
\end{align}
where the projector $\hat{P}_l(t)$ is defined as
\begin{equation}
  \label{eq:projector}
  \hat{P}_l(t) = \ket{\varphi_l^{(p)}(t)}\bra{\varphi_l^{(p)}(t)}.
\end{equation}
We will call $\ket{\varphi_n^{(p)}(t)}$ ``projected dynamical
states'', as they result from calculations that account for all
electron dynamics including strong-field ionization, but the
outcomes of these calculations are projected onto the field-free
bound-state content. Note that, evaluating the projected dynamical
states, the projected dressed ground state is only normalized without
any orthogonalization, the first excited dressed state is forced to be
orthogonal to the dressed ground state and so on.

\section{Numerical results}
\label{numerics}
Unless stated otherwise, we use \emph{atomic units} (at.~u.) in this section:
$\hbar = e = m_e = 1$, where the units of energy and 
length are $1\ \mbox{Hartree}=27.21\ \mbox{eV}$
and the Bohr radius ($5.29 \times 10^{-11}\ \mbox{m}$), respectively. 
One atomic unit of the electric field
is $5.412 \times 10^{11}\ \mbox{V}/\mbox{m}$.
To compare the different approaches to defining dressed bound states
and to illustrate the power of our analytical theory, we solve the
TDSE numerically for one electron in one spatial dimension with a
soft-core model potential:
\begin{equation}
  \label{eq:potential}
  V(z) = - \frac{1}{\sqrt{z^2 + a_\mathrm{SC}^2}}.
\end{equation}
With the soft-core parameter being equal to $a_\mathrm{SC}=0.3$, the
deepest two energy levels have energies $\epsilon_0^{(0)} = -1.75 =
-47.49\ \mbox{eV}$ and $\epsilon_1^{(0)} = -0.41 = -11.11\
\mbox{eV}$. The TDSE was solved on a grid with a step of $\Delta
z=0.1$ atomic units in a box as large as $819.2$ atomic units. We used
the first five bound states to investigate various definitions of
dressed states. The most excited of these unperturbed states has a
binding energy of $\epsilon_4^{(0)} = -0.077 = -2.10\ \mbox{eV}$.

We define the light pulses via their vector potentials: $A(t) =
-A_\mathrm{max} \cos^2\left(t/T\right) \sin(\omega t)$ for $|t| < \pi
T/ 2$ and $A(t)=0$ for $|t| \ge \pi T/ 2$.  Here, $A_\mathrm{max} =
E_\mathrm{max}/\omega$ is the peak value reached by the envelope of
the vector potential, $E_\mathrm{max}$ is the peak value reached by
the envelope of the electric field, $\omega$ is the central frequency,
and $T$ is related to the full width at half maximum (FWHM) of the
pulse intensity as $T = \mbox{FWHM}/\left[2
  \arccos\left(2^{-1/4}\right)\right]$. Given a vector potential
$A(t)$, the electric field is evaluated as $F(t)=-A'(t)$.  In our
simulations, we used a 3.5-femtosecond laser pulse
($T_\mathrm{L}=126.78=3\ \mbox{fs}$) and a 300-attosecond XUV probe pulse
($T_\mathrm{probe}=10.84=0.26\ \mbox{fs}$). The central wavelength of the laser pulse
was set to 760 nanometers ($\omega_\mathrm{L}=0.06=2.48\ \mbox{fs}^{-1}$). The central
frequency of the XUV pulse was chosen to be resonant with the
transition between the first two stationary states:
$\omega_\mathrm{probe}=\epsilon_1^{(0)} - \epsilon_0^{(0)} = 1.34 =
36.38\ \mbox{eV}$. Although we make no attempt to model a realistic
atom, the excitation from $\ket{0}$ to $\ket{1}$ may be viewed as
analogous to a dipole-allowed transition from an inner-shell orbital
of a multi-electron atom to an unoccupied orbital.

\begin{figure}[htbp]
  \centering
  \psfrag{Fig1a}[][c]{\bfseries a)}
  \psfrag{Fig1b}[][c]{\bfseries b)}
  \psfrag{Fig1xlabel}[][c]{$t$ (fs)}
  \psfrag{Fig1y2label}[][c]{$|E_\mathrm{L}(t)|$ (atomic units)}
  \psfrag{Fig1key0}[][c]{\footnotesize $\left| a_1^{(u)}(t) \right|^2$}
  \psfrag{Fig1key1}[][c]{\footnotesize $\left| a_1^{(a)}(t) \right|^2$}
  \psfrag{Fig1key2}[][c]{\footnotesize $\left| a_1^{(d)}(t) \right|^2$}
  \psfrag{Fig1key3}[][c]{\footnotesize $\left| a_1^{(p)}(t) \right|^2$}
  \includegraphics{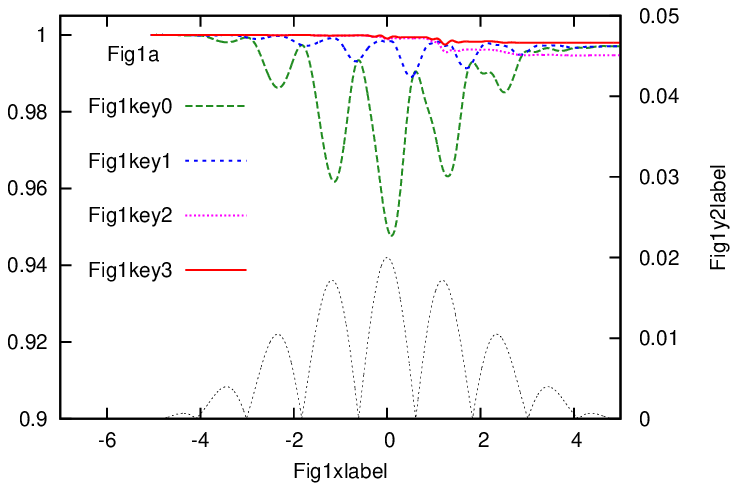}
  \includegraphics{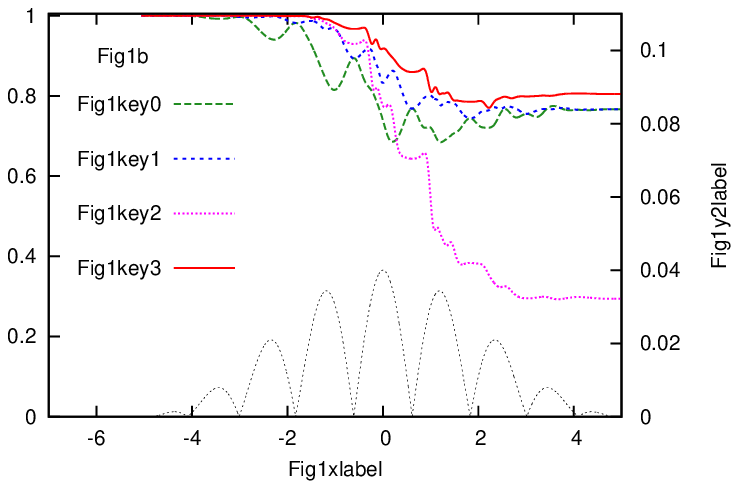}
  \caption{The probabilities of being in variously defined dressed
    states $\ket{\varphi_1(t)}$, provided that the quantum system is
    initially prepared in state $\ket{1}$. The modulus of the laser
    field is shown with black dots. The two panels correspond to
    different peak intensities $I_\mathrm{max}$ of the laser pulse: a)
    $E_\mathrm{max}=0.02\ \mbox{atomic units}\
    \left(I_\mathrm{max}=1.4 \times 10^{13}\
      \mbox{W}/\mbox{cm}^2\right)$ and b) $E_\mathrm{max}=0.04\
    \mbox{atomic units}\ \left(I_\mathrm{max}=5.6 \times 10^{13}\
      \mbox{W}/\mbox{cm}^2\right)$.}
  \label{fig:Figure1}
\end{figure}
In Fig.~\ref{fig:Figure1}, we compare the probabilities
$|a_n(t)|^2=|\braket{\varphi_n(t)}{\psi_\mathrm{L}(t)}|^2$ evaluated
using the four different definitions of dressed bound states
introduced in the previous section. In these simulations, we set
$n=1$, that is, each simulation begins with the electron being in first excited
state $\ket{1}$. For the peak electric field $E_\mathrm{max}=0.02=10^{10}\ \mbox{V}/\mbox{m}$,
the ionization probability is as little as $0.3$\%. If we knew the
perfect dressed bound states, then, for such a small ionization
probability, our model system would largely remain in
$\ket{\varphi_1(t)}$, so that $|a_1(t)|^2$ would only take values
between $|a_1(t_0)|^2=1$ and $|a_1(t_\mathrm{f})|^2=0.997$, decreasing
more or less monotonously as the initial state is slightly
depleted. Fig.~\ref{fig:Figure1} illustrates the well-known fact that
the unperturbed bound states are very far from this ideal:
$|a_1^{(u)}(t)|^2$ reaches values below $0.95$. This happens mainly
because the initial state is distorted by the external field, which
reduces the value of $|\braket{n}{\psi_\mathrm{L}(t)}|^2$. So, even though
$|a_1^{(u)}(t)|^2$ can be interpreted as the probability of being in
the unperturbed stationary state, it should not be given the
\emph{physical} meaning of being in some distorted but bound state
that, after the interaction, turns into the initial unperturbed
stationary state.

In this respect, dressed state $\ket{\varphi_1^{(a)}(t)}$, defined
by Eq.~\eqref{eq:varphi1}, performs much better, but it is not perfect
either. Even though $\ket{\varphi_1^{(a)}(t)}$ accounts for the
distortion of the initial state in the external field,
$|a_1^{(a)}(t)|^2$ in Fig.~\ref{fig:Figure1}a still has pronounced
minima at times where $E_\mathrm{L}(t)$ has its zero crossings, and
$|a_1^{(a)}(t)|^2$ reaches values that are significantly smaller than
the final probability of being in state $\ket{1}$. The main
reason for this is that $\ket{\varphi_1^{(a)}(t)}$ is a stationary
solution, while the average velocity of a bound electron can become
relatively big in an intense laser pulse. As this velocity has maxima
at zero-crossings of the electric field,
$\left|\braket{\varphi_1^{(a)}(t)}{\psi_\mathrm{L}(t)}\right|^2$ has
minima at these times.  This can also be interpreted as a breakdown of
the adiabatic (quasistatic) approximation
\cite{Christov_OE_2000}. Also the small delays between the extrema of
$E_\mathrm{L}(t)$ and the minima of $|a_1^{(u)}(t)|^2$, visible in
Fig.~\ref{fig:Figure1}a, point out some non-adiabaticity of the
electron response.

The intuitively expected step-like decrease of $|a_1(t)|^2$, as the laser field
depopulates the initial state, is best reproduced by
the dynamically dressed states $\ket{\varphi_1^{(d)}(t)}$ and the projected dynamical 
states $\ket{\varphi_1^{(p)}(t)}$, defined by
Eqs.~\eqref{eq:varphi2} and \eqref{eq:varphi3}, respectively. However,
unlike $\ket{\varphi_n^{(u)}(t)}$ and $\ket{\varphi_n^{(a)}(t)}$, these dressed
states do not necessarily turn into $\ket{n}$ once the laser pulse is gone.
In other words, $|a_1^{(d,p)}(t_\mathrm{f})|^2 \ne
|\braket{1}{\psi_\mathrm{L}(t_\mathrm{f})}|^2$ at a final time
$t_\mathrm{f}$. As long as the discrepancy is small, it is legitimate to
interpret $|a_1(t)|^2$ as the probability of being in the first excited state
dressed by the laser field. In these simulations, this discrepancy is much
smaller for $|a_1^{(p)}(t)|^2$, as compared to $|a_1^{(d)}(t)|^2$, which is
especially easy to see in Fig.~\ref{fig:Figure1}b.

\begin{figure}[htbp]
  \centering
  \psfrag{Fig2a}[][c]{\bfseries a)}
  \psfrag{Fig2b}[][c]{\bfseries b)}
  \psfrag{Fig2xlabel}[][c]{$\tau_\mathrm{probe}$ (fs)}
  \psfrag{Fig2ylabel}[][c]{$p_{0 1}[E_\mathrm{L}] /
    p_{0 1}^\mathrm{TDSE}[E_\mathrm{L} \equiv 0]$}
  \psfrag{Fig2key0}[][r]{\footnotesize $p_{0 1}^{(u)}$\hspace{3mm}}
  \psfrag{Fig2key1}[][r]{\footnotesize $p_{0 1}^{(a)}$\hspace{3mm}}
  \psfrag{Fig2key2}[][r]{\footnotesize $p_{0 1}^{(d)}$\hspace{3mm}}
  \psfrag{Fig2key3}[][r]{\footnotesize $p_{0 1}^{(p)}$\hspace{3mm}}
  \psfrag{Fig2key4}[][r]{\footnotesize $p_{0 1}^{\mathrm{TDSE}}$\hspace{3mm}}
  \includegraphics{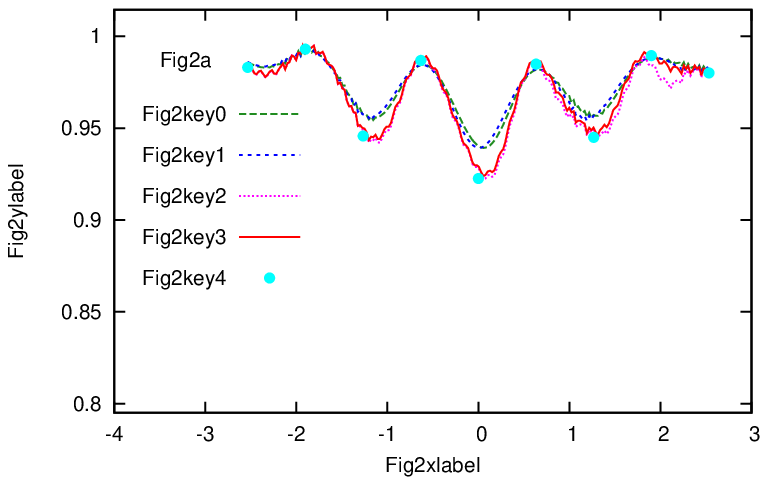}
  \includegraphics{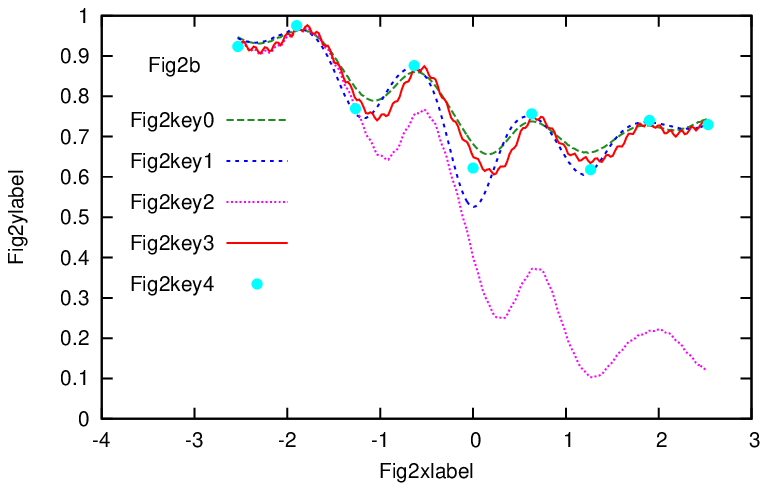}
  \caption{The effect of the laser field on the probability to find
    our model quantum system in state $\ket{0}$ at a final time
    $t_\mathrm{f}$, the initial state being $\ket{1}$. The XUV pulse,
    which is mainly responsible for the transition, arrives at a time
    $\tau_\mathrm{probe}$. The probabilities are normalized to the
    transition probability in the absence of the laser field. The
    approximate probabilities $p_{0 1}^{(u)}$, $p_{0 1}^{(a)}$, $p_{0
      1}^{(d)}$, and $p_{0 1}^{(p)}$ are evaluated with
    Eq.~\eqref{eq:transition.probability2}, using different
    definitions of dressed bound states. The exact transition
    probabilities $p_{0 1}^{\mathrm{TDSE}}$, obtained by numerically
    solving the TDSE, are shown with filled circles. The peak intensities
    of the laser pulse are the same as those in Fig.~\ref{fig:Figure1}:
    $E_\mathrm{max}=0.02$ for panel a) and $E_\mathrm{max}=0.04$ for
    panel b).}
  \label{fig:Figure2}
\end{figure}
The mere fact that, for a certain choice of dressed states,
$|a_n(t)|^2$ does not show the intuitively expected step-like behavior, does not
necessarily mean that these dressed states should not be used in our
analytical result \eqref{eq:FOP3} for the interaction with a probe
pulse. In Fig.~\ref{fig:Figure2}, we plot the results of simulations
where our model atom, initially prepared in state $\ket{1}$, interacts
with the near-infrared laser pulse $E_\mathrm{L}(t)$ and a delayed
attosecond XUV pulse $E_\mathrm{probe}(t-\tau_\mathrm{probe})$. We
plot the probability to find the atom in state $\ket{0}$ at the end of
the simulation, normalized to the probability of the same transition
in the absence of the laser pulse. Applying our model, we found that
care should be taken implementing
Eq.~\eqref{eq:transition.probability}. If a highly excited bound state
is strongly depleted by the laser pulse, $a_n(t')$ in the denominator
of Eq.~\eqref{eq:FOP3} may become very small. As a result, even small
inaccuracies in the approximations that we made deriving
Eq.~\eqref{eq:FOP3} may result in big errors. To avoid this, we restrict
the number of states used in Eq.~\eqref{eq:transition.probability}:
\begin{equation}
  \label{eq:transition.probability2}
  p_{f i} = \left| \sum_{n=0}^{\max\{i,f\}} \alpha_{n i}(t_\mathrm{f})
    \braket{f}{\varphi_n(t_\mathrm{f})} \right|^2.
\end{equation}

The filled circles in Figs.~\ref{fig:Figure2}a and \ref{fig:Figure2}b
represent the results of solving the TDSE on a grid. Comparing these
``exact'' transition probabilities with the predictions of our
analytical model, we clearly see that projected dressed states
$\ket{\varphi_n^{(p)}(t)}$, defined by Eq.~\eqref{eq:varphi3}, give
the most accurate results. At the same time, the unperturbed states
$\ket{\varphi_n^{(u)}}=\ket{n}$, which are our most primitive dressed
states, also perform surprisingly well. For the calculations with the
higher intensity, presented in Fig.~\ref{fig:Figure2}b, these states
even outperform both the adiabatic states $\ket{\varphi_n^{(a)}(t)}$ and
the dynamically dressed states $\ket{\varphi_n^{(d)}(t)}$ 
in terms of the agreement with the TDSE
calculations. For this high intensity, the biggest discrepancy between
the exact and analytical results is found for the dynamically dressed states
$\ket{\varphi_n^{(d)}(t)}$.

\begin{figure}[htbp]
  \centering
  \psfrag{Fig3a}[][c]{\bfseries a)}
  \psfrag{Fig3b}[][c]{\bfseries b)}
  \psfrag{Fig3xlabel}[][c]{$\tau_\mathrm{probe}$ (fs)}
  \psfrag{Fig3ylabel}[][c]{$p_{1 0}[E_\mathrm{L}] /
    p_{1 0}^\mathrm{TDSE}[E_\mathrm{L} \equiv 0]$}
  \psfrag{Fig3key0}[][r]{\footnotesize $p_{1 0}^{(u)}$\hspace{3mm}}
  \psfrag{Fig3key1}[][r]{\footnotesize $p_{1 0}^{(a)}$\hspace{3mm}}
  \psfrag{Fig3key2}[][r]{\footnotesize $p_{1 0}^{(d)}$\hspace{3mm}}
  \psfrag{Fig3key3}[][r]{\footnotesize $p_{1 0}^{(p)}$\hspace{3mm}}
  \psfrag{Fig3key4}[][r]{\footnotesize $p_{1 0}^{\mathrm{TDSE}}$\hspace{3mm}}
  \includegraphics{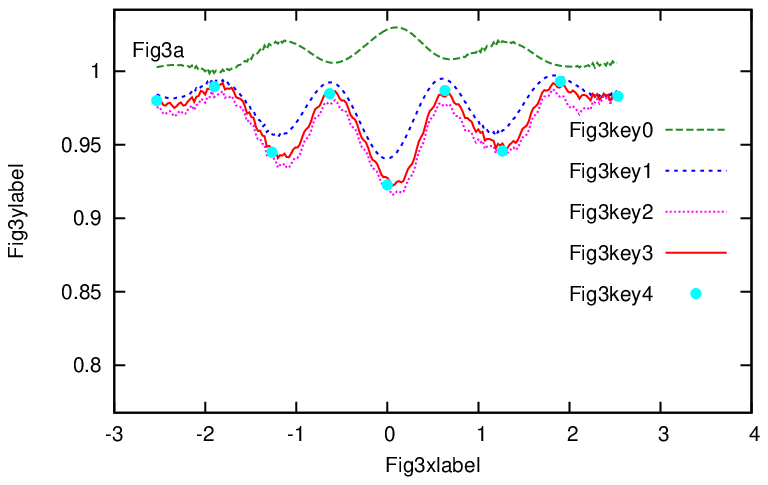}
  \includegraphics{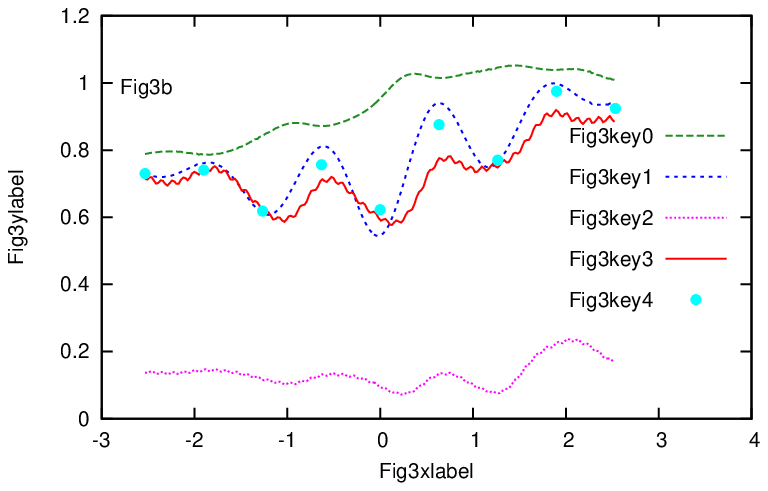}
  \caption{The effect of the laser field on the transition probability from
    state $\ket{0}$ to state $\ket{1}$. For details, see the caption of
    Fig.~\ref{fig:Figure2} and the text.}
  \label{fig:Figure3}
\end{figure}
In Fig.~\ref{fig:Figure3}, we present similar calculations for a case
more relevant to possible experiments---in this case, the XUV pulse
excites our model atom from its ground state. The figure depicts the
probability to find the atom in state $\ket{1}$ after the interaction
with the XUV and laser pulses. In contrast to Fig.~\ref{fig:Figure2},
the unperturbed states $\ket{\varphi_1^{(u)}(t)}$ fail to
describe the modulation of the transition probability by the laser
field.  Both $\ket{\varphi_1^{(a)}(t)}$ and $\ket{\varphi_1^{(p)}(t)}$
yield transition probabilities that agree with the TDSE calculations.
The corresponding discrepancies are comparable for the more intense
laser pulse (Fig.~\ref{fig:Figure3}b), but using $\ket{\varphi_1^{(p)}(t)}$
clearly gives more accurate results if the depletion of state
$\ket{1}$ is weak (Fig.~\ref{fig:Figure3}a). Thus,
in most cases, the projected dynamical states
$\ket{\varphi_n^{(p)}(t)}$, defined by Eq.~\eqref{eq:varphi3}, best
satisfy the requirements that we placed on dressed bound states.

\begin{figure}[htbp]
  \centering
  \psfrag{Fig4xlabel}[][c]{$t$ (fs)}
  \psfrag{Fig4ylabel}[][c]{$|Z_{1 0}(t)|^2$ (atomic units)}
  \psfrag{Fig4key1}[][l]{\footnotesize $E_\mathrm{max}=0.02$}
  \psfrag{Fig4key2}[][l]{\footnotesize $E_\mathrm{max}=0.03$}
  \psfrag{Fig4key3}[][l]{\footnotesize $E_\mathrm{max}=0.04$}
  \includegraphics{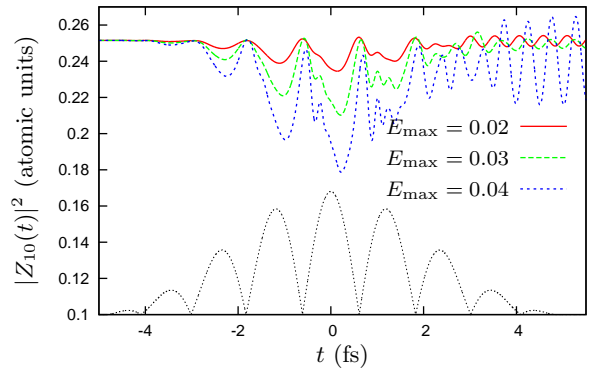}
  \caption{The squared modulus of the dipole transition matrix element
    between dressed states $\ket{\varphi_0^{(p)}(t)}$ and
    $\ket{\varphi_1^{(p)}(t)}$, as defined by Eq.~\eqref{eq:dipole}.
    $E_\mathrm{max}$ is the peak value of the electric field of the
    laser pulse.  The modulus of the laser field is schematically
    shown with black dots.}
  \label{fig:Figure4}
\end{figure}
Having identified the most promising definition of dressed bound
states, we can interpret the effect of the laser pulse on the
probabilities of XUV-induced transitions. The overall decrease of
$p_{0 1}$ with $\tau_\mathrm{probe}$ in Fig.~\ref{fig:Figure2}b, as
well as the increase of $p_{1 0}$ in Fig.~\ref{fig:Figure3}b are
clearly due to the depletion of the first excited state by the laser
field. The periodic modulations present in all the figures deserve
more consideration. In principle, they may be a consequence of two
effects: the Stark shift, which detunes the transition frequency away
from the central frequency of the XUV pulse, and the modulation of the
dressed transition matrix element \eqref{eq:dipole}. In the present
example, we find that the dominant contribution is due to the effect
of the laser field on the dressed matrix elements. To see this, we
plot $|Z_{1 0}(t)|^2$, evaluated with the aid of
$\ket{\varphi_n^{(p)}(t)}$, for different intensities of the laser
field: Fig.~\ref{fig:Figure4}. For the central half-cycles of the
laser pulse, the modulation depths of $|Z_{1 0}(t)|^2$ are very close
to those of transition probabilities in Figs.~\ref{fig:Figure2} and
\ref{fig:Figure3}.  A strong field tends to reduce $|Z_{1 0}(t)|^2$
because it displaces the electron wavefunction in the first excited
state, while the ground state in the present example has a very small
polarizability. As a consequence, the overlap between the two dressed
states decreases, which results in a smaller value of the dipole
transition matrix element between them.

The relatively fast oscillations of $|Z_{1 0}(t)|^2$ at the tail of the laser
pulse are mainly due to the multiphoton excitation of state $\ket{3}$ from
state $\ket{1}$. The energy difference between these two states is
$\epsilon_3^{(0)}-\epsilon_1^{(0)}=-0.113+0.408 \approx 0.29$ atomic units;
since the transitions from both these states to $\ket{0}$ are dipole-allowed,
this leads to quantum beats with a period of $2 \pi / 0.29 = 21\ \mbox{at.\
  u.} = 0.5\ \mbox{fs}$. These quantum beats do not, however, appear in
the probabilities of the XUV-induced transitions presented in
Figs.~\ref{fig:Figure2} and \ref{fig:Figure3} because the XUV pulse,
which is resonant with the transition between states $\ket{0}$ and $\ket{1}$,
does not have enough bandwidth to drive transitions between states $\ket{0}$
and $\ket{3}$. This fact is successfully accounted for by our analytical
theory.

\section{Conclusions and outlook}
\label{conclusions}
We have theoretically investigated how an ionizing few-cycle near-infrared
laser pulse affects single-photon bound-bound transitions driven by an
attosecond pulse of extreme ultraviolet radiation. Our approach is based on
the assumption that even when the laser field is strong enough to
significantly ionize a quantum system, one can describe bound electrons with
some \emph{dressed bound states}, which we introduce in an abstract way by a
set of simple requirements listed at the beginning of
section~\ref{general-formalism}. With such dressed states, we have obtained
relatively simple analytical results for the perturbative interaction with a
probe pulse, such as Eq.~\eqref{eq:FOP3}. For a particular model problem, we
have systematically compared four different kinds of dressed bound states and
found that the most accurate results are generally obtained with the projected
dynamical states \eqref{eq:varphi3}. These states are easily evaluated from
numerical simulations of the interaction with the laser pulse, and they make
no assumptions about the adiabaticity of the response to the laser pulse.

As the main observable in our numerical examples,
we have chosen the probability to find a single-electron atom
in a certain state after the interaction with light pulses. This was
done to test our analytical theory in a possibly simple and convincing
way. Experimentally, it is easier to measure the transient absorption
of an attosecond pulse in the presence of an ionizing laser field
\cite{Goulielmakis_Nature_2010}. It should be straightforward to
employ dressed bound states to theoretically investigate attosecond
transient absorption measurements. In this paper, we have only made a
first step in this direction by writing an explicit expression for
the dipole response associated with an XUV-driven laser-dressed
bound-bound transition: Eq.~\eqref{eq:dipole-response}.


The main purpose of this work has been to define general properties
for dressed states in strong laser-atom interactions and to explore their
existence. We have shown that quantum states with properties
close to those of idealized dressed bound states can be found.
From the fact that, overall, the projected dynamical states
$\varphi^{(p)}$ perform best, while all other states show severe
shortcomings in various situations, we can draw the following
preliminary conclusions: Not surprisingly, beyond the lowest
intensities, the unperturbed states $\varphi^{(u)}$ are not suitable
for the analysis of laser induced dynamics. Adiabatic states
$\varphi^{(a)}$ provide a good qualitative picture, but they fail when
the velocity acquired by a bound electron in a strong field becomes
significant, which is the case for systems with a reasonable
polarizability.  The dynamically dressed states $\varphi^{(d)}$ showed
us that admitting dynamics, but restricting it to the field-free bound
states, may worsen rather than improve the agreement with the full
TDSE: it appears that the restriction of the dynamics actually
introduces serious artifacts, even when comparing to the simpler
adiabatic states computed in the same subspace of field-free bound
states.  Finally, from the projected dynamical states $\varphi^{(p)}$
we learn that the field-free continuous part of the spectrum
influences the dynamics, but does not play a crucial role for the
computation of transition matrix elements between bound states.  These
four examples are clearly not the only possible choices. For example,
rather than restricting to a few field-free bound states, one may use
adiabatic or dynamic states confined to a box, which would admit
virtual continuum states without the need for a full solution of the
TDSE, as in the projected dynamical states.

We have thus shown that also in situations where significant and
strongly time-dependent ionization occurs, dressed bound states allow for a
simple analysis of experimentally observable dynamics and carry our
picture of state-depletion and perturbative transitions between the
dressed states into extreme time scales and intensity regimes. This
finding is conceptually important.  Pragmatic considerations how to
further improve the dressed states and possibly find a compromise
between accuracy and computational simplicity are left to
future work.

\section{Acknowledgments}
The authors acknowledge illuminating discussions with J.~Gagnon,
E.~Goulielmakis, and F.~Krausz.
This work was supported by the DFG Cluster of Excellence: Munich
Centre for Advanced Photonics.





\bibliography{dressed_states.bib}

\end{document}